\documentclass[aps,prc,preprint,groupedaddress,floatfix,showpacs]{revtex4}
\usepackage{graphicx}
\usepackage{dcolumn}
\usepackage{bm}

\begin{document}

\title{The Role of Surface Entropy in Statistical Emission of Massive Fragments from
Equilibrated Nuclear Systems}

\author{Jan T\~oke}
\author{Jun Lu}
\author{W.Udo Schr\"oder}
\affiliation{Department of Chemistry, University of Rochester,
Rochester, New York 14627}

\date{\today}

\begin{abstract}
Statistical fragment emission from excited nuclear systems is
studied within the framework of a schematic Fermi-gas model
combined with Weisskopf's detailed balance approach. The formalism
considers thermal expansion of finite nuclear systems and pays
special attention to the role of the diffuse surface region in the
decay of hot equilibrated systems. It is found that with
increasing excitation energy, effects of surface entropy lead to a
systematic and significant reduction of effective emission
barriers for fragments  and, eventually, to the vanishing of these
barriers. The formalism provides a natural explanation for the
occurrence of negative nuclear heat capacities reported in the
literature. It also accounts for the observed linearity of
pseudo-Arrhenius plots of the logarithm of the fragment emission
probability {\it versus} the inverse square-root of the excitation
energy, but does not predict true Arrhenius behavior of these
emission probabilities.

\end{abstract}

\pacs{21.65+f, 21.60.Ev, 25.70.Pq}

\maketitle

\section{Introduction}

Over the last decade, a considerable effort has been made to
understand the phenomenon of production of multiple
intermediate-mass fragments (IMF) in individual nuclear reaction
events. Yet, in spite of this effort, understanding of this
process is still far from being satisfactory. For example, a
consensus is still lacking on even such a fundamental issue of
whether the fragment production is dominantly a statistical or a
dynamical process. To a certain extent, such lack of consensus is
due to the fact that different models, promoting conflicting
production scenarios, appear equally successful in describing
select subsets of experimental observations.

The present study is driven by the realization that statistical
models that have been used in the interpretation of nuclear
multifragmentation data \cite{smm,mmmc,percolation,fisher,eesm}
fail to provide a coherent or satisfactory IMF production
scenario. Among other things, they fail to overcome the
fundamental difficulty of traditional equilibrium-statistical
decay models \cite{pace,gemini} in explaining how IMF production
can effectively compete with nucleon and light-charged particle
production, when the IMF emission barriers are many tens of MeV
high, compared to $\approx$ 6 MeV for protons. It was the
inability of these statistical decay models to explain the IMF
production rates that gave rise to the attractive speculation that
multiple IMF production could be a manifestation of the liquid-gas
phase transition in finite nuclear matter. This is so, because one
would expect a gas of nucleons enclosed in a container of
sufficiently large size to equilibrate under certain circumstances
into a mix of nucleons, light-charged particles, and IMFs.  This
would be a process essentially analogous to condensation of a gas
into liquid droplets. If this were the case, the abundance of IMFs
could be large and, perhaps, comparable to that observed
experimentally.

For many years, speculations regarding the relevance of the
liquid-gas transition for what is termed nuclear
multifragmentation have driven studies of this latter phenomenon.
So far, however, scrutiny of practical models featuring such a
phase transition \cite{smm,mmmc,percolation,fisher} has revealed
significant gaps in the chain of reasoning leading to the
``prediction'' of an experimentally observable liquid-gas phase
transition. For example, models based on the concept of an
existence of a freeze-out configuration \cite{smm,mmmc} utilize in
their formalisms a ``thought'' container with ideally reflecting
walls, the boundary of the freeze-out volume. Reflection of any
outgoing flux of nucleons and light-charged particles (LCP) back
into the freeze-out volume, \cite{toke_catania} is crucial for the
models allowing particles to eventually form bigger clusters, the
intermediate-mass fragments. However, since there is no known
physical boundary defining such a freeze-out volume, or mechanism
by which IMFs could pass the high saddle of emission barriers, the
predicted IMF abundance must be considered an artifact of the
modelling itself.

Similarly severe are shortcomings of approaches utilizing
percolation \cite{percolation} or Fisher's model, \cite{fisher}
which have recently been applied \cite{mor5} to nuclear
multifragmentation. \cite{schmeltzer} Notably, these models do not
even address the issue of the high Coulomb barrier inhibiting
fragment emission. The presence of a high barrier is obvious even
in the vicinity of the critical point, where the expected matter
density (in the range of 1/3 -- 1/2 of ground-state matter
density) is still high enough to trap any fragment formed inside
the expanded nuclear system in very much the same way as it would
trap fragments formed at ground state density. In the case of
Fisher's model, largely overlooked remains the fact that, by
design, the formalism is limited to events in which there is no
more than one cluster present at any time, with the balance of the
system being always in a state of a homogenous, monomeric gas of
individual molecules. While such ``either-zero-or-one-fragment''
events may be essential in a condensation theory, such as
represented by Fisher's model, they are of little, if any,
interest in the context of nuclear multifragmentation.

A somewhat different statistical approach to explaining high
yields of IMFs observed in energetic nuclear reactions was adopted
in the expanding-emitting source model (EESM). \cite{eesm} This
model considers dynamic, isentropic, self-similar expansion of a
nuclear system to densities below the equilibrium density proper
for the given total excitation energy, where it finds enhanced
fragment emission probabilities. However, as pointed out recently,
\cite{toke_catania} the EESM formalism unjustifiably derives most
of the (collective) energy necessary for a fragment to overcome
the Coulomb emission barrier, not from a thermal heat bath, as
prescribed by Weisskopf's approach, \cite{weisskopf} but from a
thermally isolated single, compressional degree of freedom.

The above discussion emphasizes the need for an alternative
scenario of statistical IMF emission. The present study
demonstrates that a simple scenario, closely related to that known
from fission studies, offers an explanation of how at moderately
high excitation energies, IMF emission can compete effectively
with nucleon evaporation. It represents an extension of ideas and
formalism presented in a recent publication. \cite{toke_surface}
Central to this formalism is the notion of a relatively high
entropy associated with the diffuse nuclear surface region (as
opposed to bulk matter). Note that it is also the increased
surface entropy that enhances the fission probability of a heavy
nucleus relative to nucleon evaporation. This latter enhancement
is commonly described in terms of the ratio $a_f/a_n$ of level
density parameters for saddle and ground-state configurations,
respectively.

The formalism adopted in the present study is described in detail
in Section~\ref{sec:formalism}. While this formalism is somewhat
schematic, it is believed to capture the essential physics
underlying the processes involved. One benefit of such a schematic
treatment is that it provides a direct insight into the phenomena
of interest, disregarding a multitude of secondary details
demanded by a more rigorous approach. Results of calculations are
presented in Section~\ref{sec:results}, while the conclusions are
presented in Section~\ref{sec:conclusions}.

\section{Theoretical Formalism}
\label{sec:formalism}

The present study assumes that an excited nuclear system expands
in a self-similar fashion so as to reach a state of approximate
thermodynamical equilibrium, where the entropy $S$ is maximal for
the given total excitation energy $E^*_{tot}$, i.e.,

\begin{equation}
{\partial S(E^*_{tot},\rho)\over \partial \rho}|_{E^*_{tot}}=0.
\label{eq:dyn_eq}
\end{equation}

The functional dependence of entropy on $E^*_{tot}$ and the bulk
nuclear matter density $\rho$ is evaluated using the Fermi-gas
model relationship

\begin{equation}
S=2\sqrt{aE^*_{th}}=2\sqrt{a(E^*_{tot}-E_{compr})},
\label{eq:entropy}
\end{equation}

\noindent where $a$ is the level density parameter, $E^*_{th}$ is
the thermal excitation energy and $E_{compr}$ is the collective
compressional energy. The dependence of $S$ on bulk nuclear matter
density $\rho$ arises in Eq.~\ref{eq:entropy} through the
dependence of both, the level density parameter $a$
(``little-$a$'') and the compressional energy $E_{compr}$ on the
matter density.

The dependence of little-$a$ on the nuclear matter density for
infinite nuclear matter is given by the Fermi-gas model:

\begin{equation}
a=a_o({\rho\over \rho_o})^{-{2\over 3}}, \label{eq:little_a_fg}
\end{equation}

\noindent where $a_o$ is the level density parameter for the
nuclear matter at ground-state matter density $\rho_o$.

The above equation holds approximately also for finite nuclei if
the expansion or compression of these nuclei is assumed to occur
in a self-similar fashion. This is so because for finite nuclei,
the little-$a$ parameter consists \cite{tok81,ign75} of volume and
surface terms, $a_V$ and $a_\sigma$, respectively, both of which
are proportional to $\rho^{-2/3}$ under the assumption of
self-similar expansion:

\begin{equation}
a=a_V+a_\sigma=A({\rho\over \rho_o})^{-{2\over
3}}\alpha_V+A^{2\over 3}({\rho\over \rho_o})^{-{2\over
3}}\alpha_\sigma, \label{eq:little_a_finite}
\end{equation}

\noindent where $\alpha_V$ and $\alpha_\sigma$ are volume and
surface coefficients, respectively, independent of bulk nuclear
matter density.

The term ``self-similar expansion'' is used here to describe a
type of expansion in which any change in the matter density
profile is reducible to a simple rescaling of the radial
coordinate, such that:

\begin{equation}
f_\rho(r)=c^3f_o(cr), \label{eq:self_similar}
\end{equation}

\noindent where $f_o(r)$ is the ground-state density profile
function and $c$ is a scaling constant.

The presence of a surface contribution to the level density
parameter is of crucial importance in the present study as it
describes the part $S_\sigma$ of entropy $S$ of the system
associated with the diffuse surface domain and is seen to have
significant effect on the fragment emission probability. One has

\begin{equation}
S=S_V+S_\sigma=a_V T+a_\sigma T, \label{eq:surf_entropy}
\end{equation}

\noindent where $S_V$ is the entropy of the bulk matter and T is
the system temperature.

The compressional energy in Eq.~\ref{eq:entropy} is approximated
in the present study following the schematic prescription proposed
in the expanding emitting source model EESM \cite{eesm}, i.e.,

\begin{equation}
\epsilon_{compr}=\epsilon_b(1-{\rho\over \rho_o})^2,
\label{eq:compr}
\end{equation}

\noindent where $\epsilon_{compr}$ and $\epsilon_b$ are the
compressional and the ground-state binding energy per nucleon of
the system, respectively. Note that Eq.~\ref{eq:compr} ensures
that the compressional energy varies parabolically with $\rho$,
from zero at ground-state density ($\rho_o$) to $\epsilon_b$ at
zero density.

Equations~\ref{eq:dyn_eq}--\ref{eq:little_a_fg} and \ref{eq:compr}
allow one to obtain an analytical expression for the equilibrium
density, $\rho_{eq}/\rho_o$, of nuclear matter as a function of
the excitation energy per nucleon, $\epsilon^*_{tot}=E^*_{tot}/A$,
where $A$ is the mass number of the system:

\begin{equation}
{\rho_{eq}\over \rho_o}={1\over
4}(1+\sqrt{9-8{\epsilon^*_{tot}\over \epsilon_b}}\;).
\label{eq:EquilDens}
\end{equation}

Equation~\ref{eq:EquilDens} reflects the fact that for
$\epsilon^*_{tot}< \epsilon_b$, the system is bound as far as the
self-similar expansion mode is concerned, and features a single
maximum in entropy in the range of matter densities $1/2 \leq
\rho_{eq}/\rho_o \leq 1$. For $\epsilon^*_{tot}\geq \epsilon_b$,
the system is essentially unbound with respect to the self-similar
expansion mode, with the entropy diverging as the matter density
$\rho$ tends to zero. However, for $(9/8)\epsilon_b\geq
\epsilon^*_{tot}\geq \epsilon_b$, the system still has a local
metastable maximum at a finite density given by
Eq.~\ref{eq:EquilDens}, i.e., in the range of matter densities
$1/4 \leq \rho_{eq}/\rho_o \leq 1/2$. The latter metastable (with
respect to self-similar expansion mode) maximum in entropy is
separated from the divergence at zero density by a minimum at
$\rho_{saddle}$, where

\begin{equation}
{\rho_{saddle}\over \rho_o}={1\over
4}(1-\sqrt{9-8{\epsilon^*_{tot}\over \epsilon_b}}\;).
\label{eq:SaddleDens}
\end{equation}

\noindent Here, $E_b=A\epsilon_b$.

Further, one can obtain the caloric equation of state of an
equilibrated system (i.e. at constant, zero external pressure) by
using the Fermi-gas model relationship between the system
temperature $T$ and thermal excitation energy $E^*_{th}$

\begin{equation}
T=\sqrt{{E^*_{th}\over a}}=({\rho_{eq}\over \rho_o})^{1\over
3}a_o^{-{1\over 2}}\sqrt{E^*_{tot}-E_b(1-{\rho_{eq}\over
\rho_o})^2}. \label{eq:caloric}
\end{equation}

The probability $p$ of emitting a fragment from an equilibrated
excited system as defined above can be evaluated using the
Weisskopf formalism: \cite{weisskopf}

\begin{equation}
p\propto e^{\Delta S}=e^{S_{saddle}-S_{eq}}, \label{eq:em_prob}
\end{equation}

\noindent where $S_{saddle}$ and $S_{eq}$ are saddle-point and
equilibrium-state entropies, respectively. The latter two
entropies can be calculated using Eq.~\ref{eq:entropy}:

\begin{equation}
S_{eq}=2\sqrt{a_A[E^*_{tot}-E_b(1-{\rho_{eq}\over \rho_o})^2]},
\label{eq:eq_entropy}
\end{equation}

\begin{equation}
S_{saddle}=S_{res}+S_{frag}=2\sqrt{(a_{res}+a_{frag})E^{*th}_{saddle}}.
\label{eq:saddle_entropy}
\end{equation}

In Eqs.~\ref{eq:eq_entropy} and \ref{eq:saddle_entropy}, $a_A$,
$a_{res}$ and $a_{frag}$ are the level density parameters of the
system at equilibrium, the residue, and the fragment,
respectively, while $E^{*th}_{saddle}$ is the thermal excitation
energy of the system in the saddle-point configuration. The latter
quantity is calculated as

\begin{equation}
E^{*th}_{saddle}=E^*_{tot}-E_b(1-{\rho_{eq}\over
\rho_o})^2-V_{saddle}, \label{eq:estar_saddle}
\end{equation}

\noindent where $V_{saddle}$ is the (collective) saddle-point
energy.

Formally, the emission probability $p$ can be also expressed in
terms of an effective emission barrier $B_{eff}$,

\begin{equation}
\label{eq:B_eff1} p\propto e^{-{B_{eff}\over T}},
\end{equation}

\noindent when one sets

\begin{equation}
B_{eff}=-T\Delta S \label{eq:B_eff}.
\end{equation}

A selection of results of calculations performed using the above
formalism is presented in Section~\ref{sec:results} below.

\section{Results of Model Calculations}
\label{sec:results}

Results of the calculations performed in the framework of the
formalism presented in Section~\ref{sec:formalism} are presented
in Figs.~1--7. In these calculations, values of $\alpha_V$=1/14.6
MeV$^{-1}$ and $\alpha_\sigma$=4/14.6 MeV$^{-1}$ have been assumed
for the coefficients $\alpha_V$ and $\alpha_\sigma$, as suggested
in literature \cite{tok81}. Further, $\epsilon_b$=8 MeV was
assumed \cite{eesm} for the ground-state binding energy per
nucleon, while the saddle-point collective energy was approximated
by the Coulomb energy of the residue and fragment represented by
two touching spheres of radius parameter
$r_{Coul}=1.3(\rho/\rho_o)^{-1/3}$ fm. The calculations were made
for excited $^{197}$Au nuclei.

\begin{figure}
\includegraphics{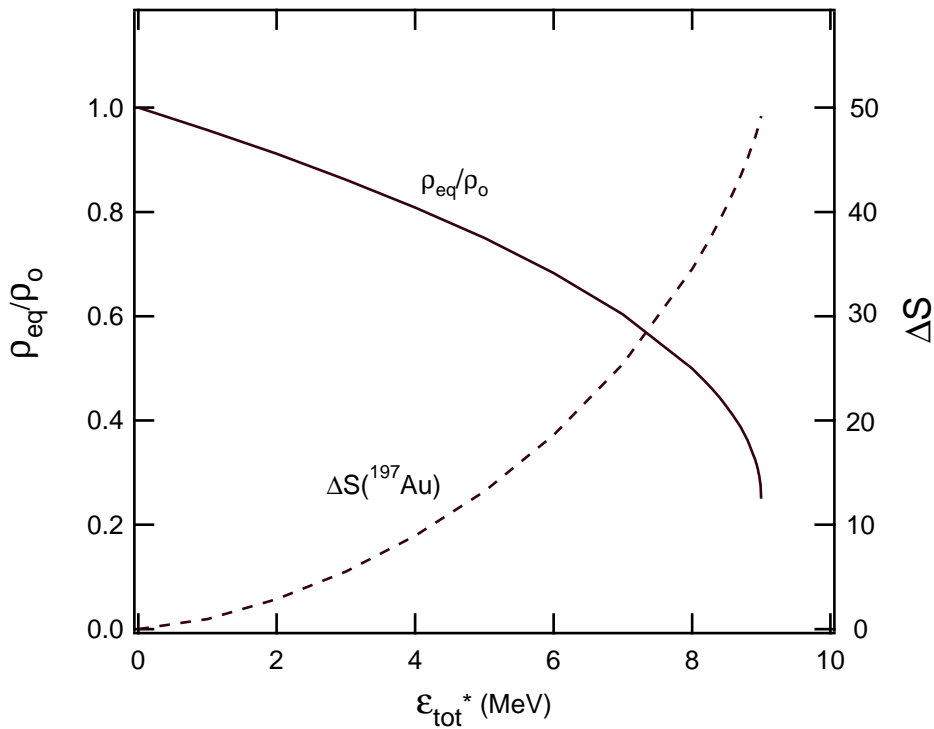}
\caption{\label{fig:equil_dens} Equilibrium density of bulk
nuclear matter (solid curve) and the net gain in entropy realized
by an $^{197}Au$ system as a result of the relaxation of the
self-similar expansion mode (dashed curve), plotted as functions
of total excitation energy per nucleon.}
\end{figure}

Fig.~\ref{fig:equil_dens} illustrates the dependence of the
equilibrium density, $\rho_{eq}$, of bulk matter on the excitation
energy per nucleon (solid curve). As seen in this figure, for the
range of excitation energies readily accessible in experiments,
the bulk matter density in a state of maximum entropy differs
substantially from that of the nuclear ground state. This affects
both, the caloric equation of state and the fragment emission
probability, two entities of considerable interest.

The dashed curve seen in Fig.~\ref{fig:equil_dens} illustrates the
net gain in entropy resulting from the relaxation of the
self-similar expansion mode in an excited $^{197}Au$ system. Large
gains in entropy associated with the relaxation of this mode
emphasize the importance of this mode for a statistical
description of excited nuclear systems, notably for models based
on the concept of microcanonical \cite{mmmc} or
pseudo-microcanonical \cite{smm} equilibrium.

\begin{figure}
\includegraphics{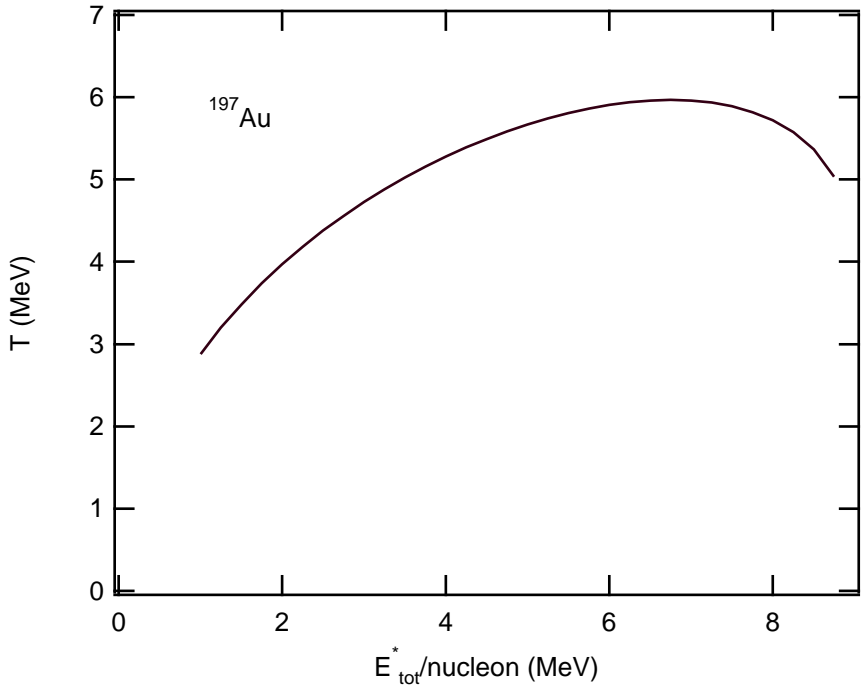}
\caption{\label{fig:caloric_curve} Caloric curve for the
$^{197}$Au system.}
\end{figure}

The caloric curve calculated for states of maximum entropy, i.e.,
for the states of equilibrium density $\rho_{eq}$ is depicted in
Fig.~\ref{fig:caloric_curve}. Not surprisingly, this curve shows
considerable deviation from the simple Fermi-gas form of $T\propto
\sqrt{E^*_{tot}}$. Note that in experiments, such as the recently
reported ISIS experiment, \cite{isis} it is $E^*_{tot}$ and not
the purely thermal contribution to it, $E^*_{th}$, that is in fact
measured. This is so, because the static compressional energy
$E_{compr}$ is experimentally undistinguishable from thermal
excitation $E^*_{th}$.

Most notably, the caloric curve seen in
Fig.~\ref{fig:caloric_curve} features a maximum temperature of
approximately $T_{max}$=6 MeV, an indication that for higher
temperatures the system is inherently unstable and does not find
an equilibrium density. In other words, under the assumption that
its matter distribution is homogeneous, a nuclear system placed in
a heat bath of $T>6MeV$ would expand indefinitely, for which it
would derive increasing amounts of energy from the heat bath. In a
more realistic case, which is beyond the present consideration,
the system would likely decay into a ``gas'' of fragments and free
nucleons that would continue its indefinite expansion.

\begin{figure}
\includegraphics{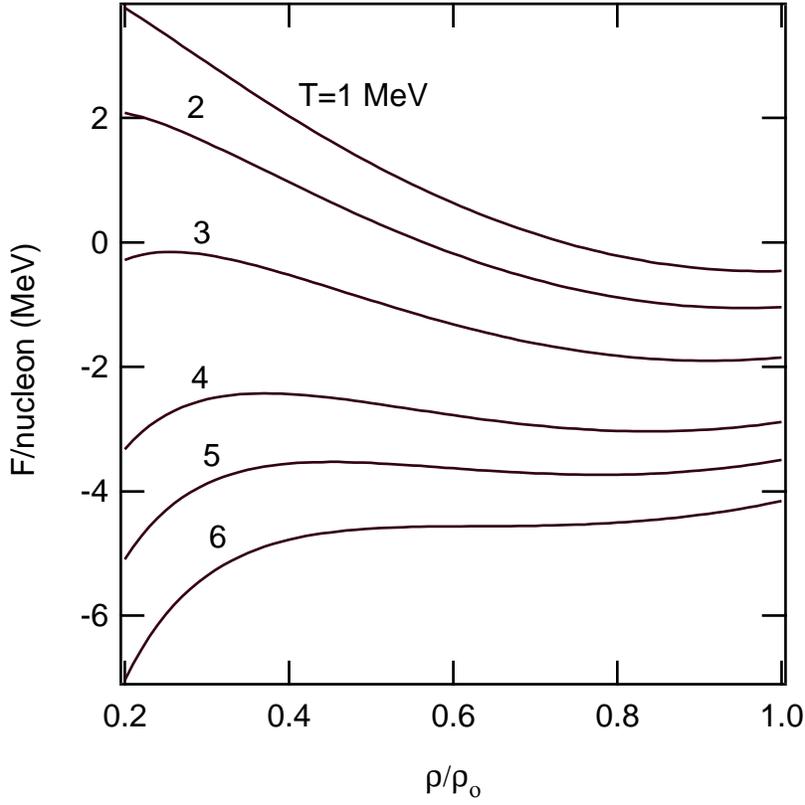}
\caption{\label{fig:free_energy} Free energy for the $^{197}$Au
system as a function of nuclear matter density and temperature.}
\end{figure}

The particular form of the caloric curve seen in
Fig.~\ref{fig:caloric_curve} can, perhaps, be better understood
when inspecting the dependence of the free energy $F$ on matter
density $\rho$ and temperature $T$, which is is illustrated in
Fig.~\ref{fig:free_energy}. The calculations show that, at
temperatures below $T\approx 6 MeV$, the free energy as a function
of $\rho$ features two equilibrium points, a local minimum and
local maximum, where the latter reflects an unstable equilibrium
These two points predicted for any temperature $T$ of less than
approximately 6 MeV, correspond to different excitation energies
per nucleon, $\epsilon^*_{tot}$. Note that at both these extremal
points, the pressure of the system at its surface is zero. This is
so, because the pressure is proportional to the partial derivative
of the free energy with respect to density at constant entropy,
and both, the free energy and the entropy are stationary with
respect to the matter density at the extremal points in question.
At $T\approx 6 MeV$, the minimum and maximum in $F$ as a function
of $\rho$ merge into an inflection point and, then, at even higher
temperatures, the free energy features only a monotonic decrease
with decreasing $\rho$. The inflection point in $F$ corresponds to
maximum in $T$ as a function of $E^*_{tot}$. Obviously, a
monotonic decrease in free energy with decreasing matter density
means that no equilibrium density exists for a given temperature.

\begin{figure}
\includegraphics{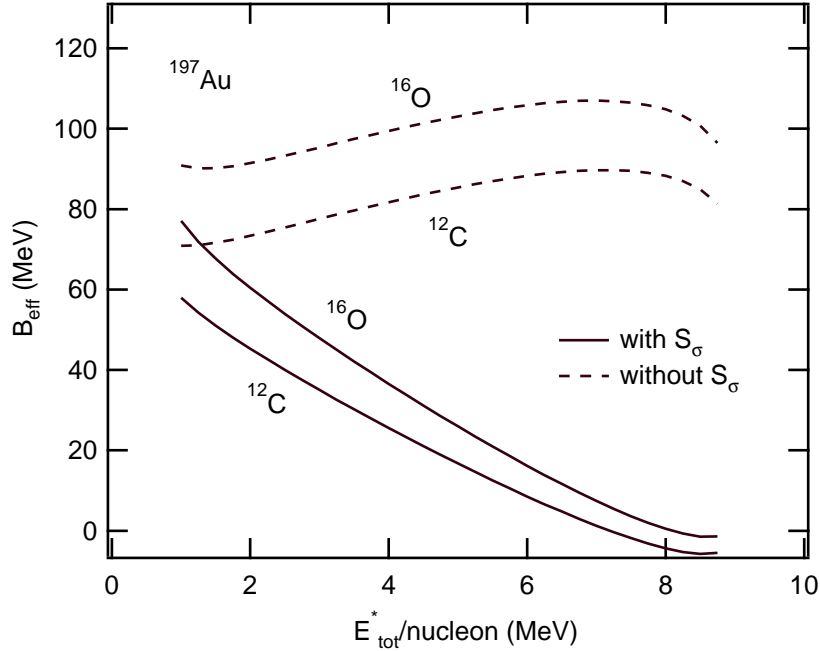}
\caption{\label{fig:B_eff} Effective barriers for the emission of
$^{12}$C and $^{16}$O fragments from equilibrated excited
$^{197}$Au systems as functions of excitation energy per nucleon.}
\end{figure}

The role of the surface entropy in fragment emission is clear from
Fig.~\ref{fig:B_eff} illustrating the dependence of effective
barriers (see Eq.~\ref{eq:B_eff}) for the emission of $^{12}$C and
$^{16}$O from exited $^{197}$Au nuclei. Solid lines in this figure
illustrate the results obtained when the surface entropy is taken
into consideration via inclusion of $a_\sigma$ in the expression
for the level density parameter $a$ (see
Eq.~\ref{eq:little_a_fg}), while dashed lines represent results of
calculations in which the surface entropy effects were neglected,
i.e., in which $a$ was assumed simply proportional to the mass
number.

It is important to note that with the reduction in the effective
barrier being dominantly due to the surface entropy effects, no
analogous reduction is expected for the emission of nucleons or
light charged particles. In the latter case, the effective
barriers are expected to show trends similar to those shown by the
dashed lines in Fig.~\ref{fig:B_eff}.

\begin{figure}
\includegraphics{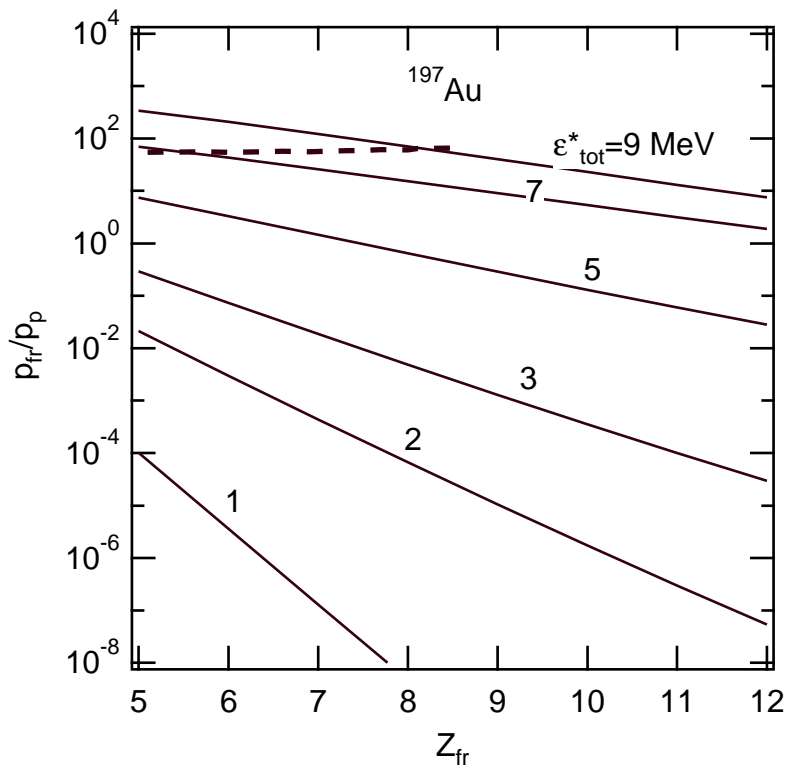}
\caption{\label{fig:z_spectra} Distributions of relative emission
probabilities of various IMFs from an excited, equilibrated
$^{197}$Au system as functions of the total excitation energy
(solid curves). The dashed line represents the boundary of the
domain of dynamical instability of the system. (See text)}
\end{figure}

The effects of the surface entropy on the relative emission rates
of various IMFs are illustrated in Fig.~\ref{fig:z_spectra},
displaying the reduced probabilities (taken as bare Boltzmann
factors) for the emission of IMFs of different atomic numbers from
an equilibrated $^{197}$Au system as functions of the total
excitation energy per nucleon. The spectra are normalized to the
probability for emitting a proton, bound to the system by $Q_p$=8
MeV and experiencing a Coulomb barrier of 4 MeV (corresponding to
a Coulomb radius parameter of $r_C$=1.3 fm). As seen in this
figure, already at excitation energies per nucleon of the order of
3 MeV/nucleon, fragment emission begins to compete effectively
with proton emission and, then, at higher excitation energies,
fragment emission becomes the dominant decay mode. Note, that the
highest probabilities seen in Fig.~\ref{fig:z_spectra}, above the
dashed line, reflect dynamical instability with respect to the
fragment emission mode. In fact, Weisskopf's approach
\cite{weisskopf} is inapplicable in the domain above the dashed
line, as in this domain, the saddle-point entropy exceeds that for
the ``equilibrium'' density of a self-similarly expanded system.
In the calculations, it was assumed that the mass number of the
fragment is twice the atomic number, i.e., $A_{fr}=2Z_{fr}$.

\begin{figure}
\includegraphics{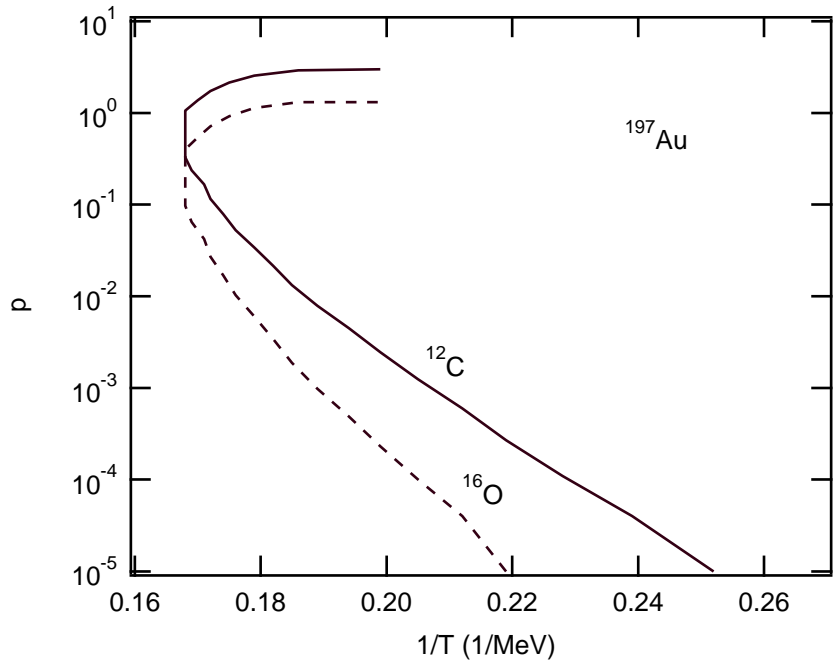}
\caption{\label{fig:true_arrhenius} Arrhenius plots for emission
of $^{12}$C and $^{16}$O fragments from excited $^{197}$Au
systems.}
\end{figure}

Since the magnitude of the effective barrier depends on the
excitation energy and, thus, on temperature $T$, and since $T$ is
a non-monotonic function of $E^*_{tot}$, one would expect the
fragment emission probabilities $p$ to deviate significantly from
a simple Arrhenius law with an exponential functional dependence
on the inverse temperature $1/T$. This expectation is confirmed by
Fig.~\ref{fig:true_arrhenius}, where logarithmic plots of $p$ {\it
versus} $1/T$ are seen to feature clear deviations from linearity,
including a prominent ``back-bending''. The latter back-bending is
obviously expected in view of the maximum in the caloric curve
seen in Fig.~\ref{fig:caloric_curve}. Such a behavior is also
suggested by the general deviation of the caloric equation of
state from the one for a low-temperature Fermi gas.

\begin{figure}
\includegraphics{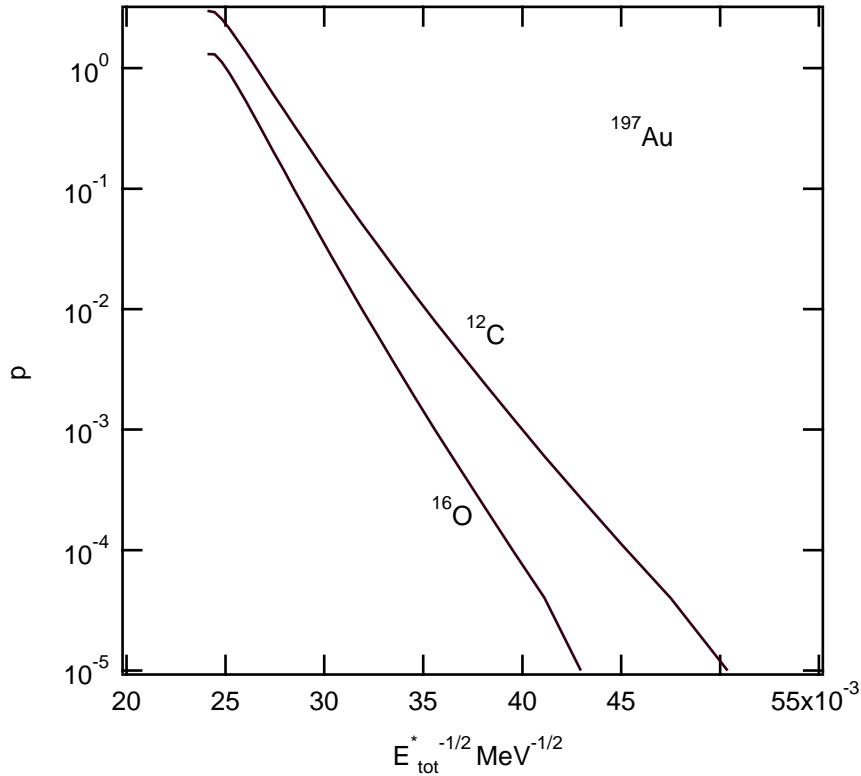}
\caption{\label{fig:pseudo_arrhenius} Pseudo-Arrhenius plots for
emission of $^{12}$C and $^{16}$O fragments from excited
$^{197}$Au systems.}
\end{figure}

On the other hand, pseudo-Arrhenius plots depicted in
Fig.~\ref{fig:pseudo_arrhenius}, where the logarithm of the
emission probability, $ln(p)$, is plotted {\it versus} the inverse
square root of the total excitation energy, $1/\sqrt{E^*_{tot}}$,
, are to a good approximation straight lines. This observation
comes as a surprise, as it cannot readily be expected based on the
details of the theoretical formalism employed. Therefore, no
simple explanation for such a linearity can be offered at this
time, the very rationale behind the construction of such
pseudo-Arrhenius plots being numerous experimental findings.
\cite{mor1,mor2,mor3,mor4,mor5} While it may be purely fortuitous,
the linear character of pseudo-Arrhenius plots predicted by the
present formalism seems to be in agreement with experimental
observations.

\section{Conclusions}
\label{sec:conclusions}

A formalism has been developed to describe quantitatively, albeit
in a schematic fashion, a scenario of purely statistical emission
of massive fragments from finite equilibrated systems. The
formalism recognizes the importance of thermal expansion of hot
matter and considers the stability of such systems at equilibrium
nuclear densities. The important role of surface entropy consists
in an effective ``softening'' of the nuclear surface, resulting in
enhanced fragment emission probabilities. The decay scenario
underlying this formalism is that of a thermally expanded system
with developed thermal fluctuations of the diffuse surface. Within
such a scenario, fragment emission is naturally viewed as
analogous to fission.

The calculations performed within the framework of the proposed
formalism show that, with increasing excitation energy and mostly
due to effects of increased surface entropy, the effective
barriers for fragment emission decrease  and, eventually, vanish.
As no equivalent decrease is expected for nucleons and light
charged particles, one expects that at high excitation energies,
fragment emission competes effectively with nucleon and LCP
emission.

The formalism presented here also predicts an occurrence of
negative heat capacity in highly excited nuclear systems. It also
predicts linear pseudo-Arrhenius plots, where the logarithm of the
fragment emission probability, $ln(p)$, is plotted {\it versus}
the inverse square-root of the excitation energy, $1/\sqrt{E^*}$.
This latter prediction is found in agreement with numerous
experimental observations. \cite{mor1,mor2,mor3,mor4} However, as
also expected, the formalism predicts non-linear true Arrhenius
plots, where the independent variable is the inverse temperature,
$1/T$, rather than $1/\sqrt{E^*}$. This is so, because the
effective emission barriers depend rather strongly on excitation
energy (and temperature) and, hence, fail to satisfy one of the
fundamental prerequisites of the Arrhenius law.

The simplicity of the presented formalism should allow one to
incorporate it into existing statistical decay codes. For purposes
where the ultimate goal is to obtain an agreement with
experimental data, it would be desirable, however, to improve the
formalism itself. For example, one should use more realistic (and
lower) saddle-point energies, instead of the schematic Coulomb
energies of touching spheres. One should also be mindful of
differences in equilibration times for nucleon and LCP emission
modes, on the one hand, and for different IMF emission and fission
modes, on the other hand. It is natural to expect that modes
requiring major rearrangement of mass require longer times to
equilibrate than modes that do not require such a rearrangement.
Consequently, one would expect all IMF emission modes to attain
equilibrium later than nucleon emission modes, and fission modes
to reach equilibrium even later. As a result, early in the decay
process, while the IMF degree of freedom is approaching
equilibrium, one would expect to see ``pre-IMF nucleons'', in much
the same way as one sees pre-fission neutrons.
\cite{hilscher,hinde} Similarly, one would expect to see
pre-fission IMF emission in situations when for the same systems
at equilibrium, fission would dominate. Importantly, the latter
pre-fission IMF emission modes may succeed in removing enough
charge and excitation energy from the system so as to pre-empt
fission altogether - a phenomenon that has recently been observed
experimentally. \cite{ben}

\newpage

\begin{acknowledgments}
This work was supported by the U.S. Department of Energy grant No.
DE-FG02-88ER40414.
\end{acknowledgments}

\end{document}